\documentstyle[twocolumn,prl,aps,epsf]{revtex}

\title{Staggered-vorticity correlations in a lightly doped $t$-$J$ model:
a variational approach.}

\author{D. A. Ivanov\cite{DI-address},
Patrick A. Lee, and Xiao-Gang Wen}

\address{Department of Physics and Center for Materials Science and
Engineering, MIT,
Cambridge, MA 02139}

\begin{document}

\twocolumn[\hsize\textwidth\columnwidth\hsize\csname @twocolumnfalse\endcsname

\maketitle

\begin{abstract}
We report staggered vorticity correlations of current in the d-wave
variational wave function for the lightly-doped $t$-$J$ model. Such
correlations are explained from the $SU(2)$ symmetry relating $d$-wave
and staggered-flux mean-field phases. The correlation functions
computed by the variational Monte Carlo method suggest that pairs
are formed of holes circulating in opposite directions.
\end{abstract}

\pacs{#1}

]

\narrowtext

Flux phases have been proposed as mean-field solutions to
two-dimensional antiferromagnets \cite{Zhang}.
For the undoped Heisenberg antiferromagnet, the staggered-flux variational
wave function gives a relatively good energy \cite{Gros,Yokoyama,Dmitriev},
even without the Neel long-range order. However, for doped
systems staggered-flux phases would normally break the translational
and the time-reversal symmetries, except at the flux value of $\pi$.
Besides, the physical meaning of the flux is not transparent. It
has been suggested that flux is related to the spin-chirality
correlations \cite{Lee-Nagaosa}, but such correlations are very complicated
for both experimental and theoretical study.

Remarkably, in the case of doped $t$-$J$ or Hubbard models, there is one more
indication of a staggered-flux phase. Namely, the current-current
correlations may show the staggered-flux pattern inherited from the
mean-field phase. We find such a pattern in the Gutzwiller-projected
$d$-wave variational wave function for the $t$-$J$ model. Those correlations
may be explained by the $SU(2)$ equivalence between the $d$-wave-pairing
and staggered-flux phases.
Finally, we interpret the staggered-flux structure of current correlations
as a pairing between holes of opposite ``staggered-vorticity''.

Consider first the $d$-wave variational wave function \cite{Gros,Yokoyama}:
\begin{equation}
\Psi=P_G \prod_k (u_k+v_k c^\dagger_{k\uparrow} c^\dagger_{k\downarrow})
|0\rangle,
\label{wf-pairing}
\end{equation}
where $P_G$ is the projection onto the states with a fixed number of
electrons and without doubly-occupied sites. The coherence factors
$u_k$ and $v_k$ are of the BCS form for $d$-wave pairing:
\begin{eqnarray}
a_k &\equiv& {v_k\over u_k}={\Delta_k\over
\xi_k+\sqrt{\xi_k^2+\Delta_k^2}},
  \nonumber\\
\xi_k &=&-2(\cos k_x + \cos k_y)-\mu,
   \\
\Delta_k &=& \Delta(\cos k_x -\cos k_y).
  \nonumber
\end{eqnarray}
For simplicity, we consider the case $\mu=0$. Tuning $\mu$ improves the
energy by only
 0.2\%, and the current
correlations reported in this paper remain
practically the same.

\begin{figure}
\epsfxsize=3.0in
\centerline{\epsffile{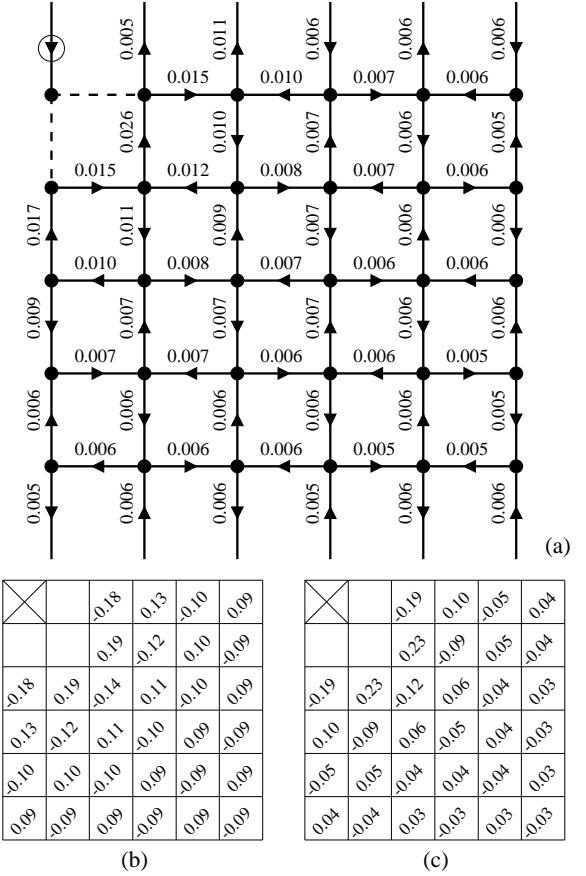}}
\smallskip
\caption{(a) Current-current correlations for 2 holes in the
$10\times 10$ lattice at $\Delta=0.55$ and $\mu=0$. Boundary conditions are
periodic in one and antiperiodic in the other direction
(the data are averaged over the two orientations).
The number on a link
is the correlation of the current on this link and of the current on the
circled link divided by hole density. The arrows point in the direction of the
positive correlations of the current.
(b) The same data in the form of vorticity correlations. The number on a
plaquet is the vorticity correlation divided by $x$ with the crossed plaquet.
(c) Same as (b) for 10 holes in $10\times 10$ lattice.
}
\label{short-fig1}
\end{figure}

Using the variational Monte Carlo (VMC) method, we computed the current-current
correlations for the wave function (\ref{wf-pairing}) in a finite system
(2 holes in the
$10\times10$ lattice
with periodic-antiperiodic boundary conditions, $\Delta=0.55$).
The current on a link $(ij)$ is defined as
\begin{equation}
j_{ij}=-j_{ji}=i(c^\dagger_{\alpha i} c_{\alpha j} -
c^\dagger_{\alpha j} c_{\alpha i}).
\label{current-def}
\end{equation}
The correlation function exhibits a staggered-flux structure
(Fig.~\ref{short-fig1}).
Since the current vanishes as the hole concentration $x \rightarrow 0$, we
have divided the correlation function by $x$.
 This behavior of the current correlations
is observed in the whole range of the gap values $\Delta$.
The correlations are smaller at small $\Delta$, which will
be interpreted as the limit of staggered flux going to zero.
We introduce {\it vorticity} $V$ for any plaquet
as a sum of the currents around it in the counterclockwise direction:
$V=(j_{12}+j_{23}+j_{34}+j_{41})_\Box$.
The vorticity correlations (Fig.~\ref{short-fig1}b) obtained
from the current-current correlations in Fig.~\ref{short-fig1}a
have alternating sign with a phase shift
of $\pi$, so that the sign of $\langle V(0) V(R)\rangle $ is
$(-1)^{R_x+R_y+1}$.

This staggered-vorticity correlation is a surprising consequence of the
projection,
because it is absent in the unprojected $d$-wave wavefunction.  In order to
understand
this, we show below that the same wavefunction can be written as a
projection of a
staggered-flux state.  This has the advantage that properties that are
obscure in one
representation may become obvious in another.  For example, the staggered-flux
wavefunction is an insulator with gap nodes.  The appearance of
superconductivity
after projection is a surprise, but vorticity and attraction between holes
appear
naturally, as we shall discuss below.

The basic starting point is the $SU(2)$ symmetry in the fermion
representation of the
$t$-$J$ model.  This is well understood in the undoped case \cite{Affleck},
where
$SU(2)$ doublets $\psi_{\uparrow i} = (f_{\uparrow i},f_{\downarrow
i}^\dagger)$ and
$\psi_{\downarrow i} = (f_{\downarrow i},-f_{\uparrow i}^\dagger)$
represent the
destruction of spin up and spin down in the subspace of one fermion per
site.  Wen and
Lee \cite{Wen-Lee} extended this symmetry away from half filling by
introducing a
doublet of bosons $b_i = (b_{1i},b_{2i})$.  The physical electron is
represented as an
$SU(2)$ singlet formed out of the fermion and boson doublets
$
c_{\alpha i} = \frac{1}{\sqrt 2} b_i^\dagger \psi_{\alpha i}
$.

At the level of variational wave functions, the constraint of no double
occupation is
enforced by projecting the fermion-boson wave function onto the
$SU(2)$-singlet subspace of the Hilbert space. On each site, there are
only three physical states: spin-up,
spin-down and a hole,
\begin{equation}
|\!\uparrow\rangle=f^\dagger_\uparrow|0\rangle, \;\;\;
|\!\downarrow\rangle=f^\dagger_\downarrow|0\rangle, \;\;\;
|\star\rangle={1\over\sqrt 2}(b^\dagger_1+b^\dagger_2
f^\dagger_\uparrow f^\dagger_\downarrow)|0\rangle.
\label{singlet-states}
\end{equation}
The projector may be written as
\begin{equation}
P_{SU(2)}=\prod_i
\left( |\!\uparrow\rangle\langle\uparrow\!| +
|\!\downarrow\rangle\langle\downarrow\!| + |\star\rangle\langle\star|
\right)_i.
\label{psu2}
\end{equation}
It should be applied to a mean-field state  with the total number of bosons
defining the number of doped holes.  The mean-field Hamiltonian has the form
\begin{eqnarray}
H= & & \sum_{\{ ij\} }\left[
J \psi_{\alpha i}^\dagger U_{ij}\psi_{\alpha j}
 + t (b_i^\dagger U_{ij} b_j + {\rm h.c.})
\right] \nonumber \\
& & +
\sum_i a_i^\mu ({1\over2}\psi_{\alpha i}^\dagger
\tau_\mu \psi_{\alpha i} + b_i^\dagger \tau_\mu b_i),
\label{mf-ham}
\end{eqnarray}
where $U_{ij}$ are $SU(2)$ matrices representing generalized hopping amplitudes
(mean-field parameters) for nearest-neighbor sites $i$ and $j$.

\begin{figure}
\epsfxsize=2.8in
\centerline{\epsffile{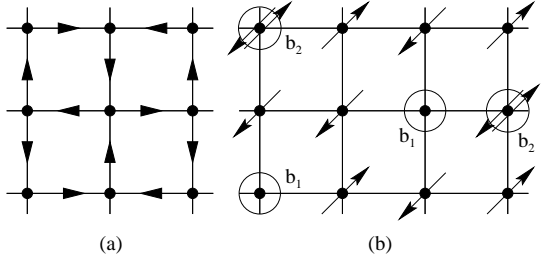}}
\smallskip
\caption{(a)~Staggered-flux phase. Links with arrows correspond to
$a_{ij}=\varphi/4$
in the direction of the arrow. (b)~A typical configuration of the
half-filled fermion
state.  Arrows denote fermions.  Circled sites are physical holes which are
spin
singlets made up of either empty or two-fermion sites.  $b_1$ and $b_2$
bosons are
assigned to these respective sites.}
\label{short-fig2}
\end{figure}

In the underdoped region, the mean field solution is  the staggered-flux phase
characterized by
\begin{equation}
U_{ij}=e^{i a_{ij} \tau_3}, \qquad a^\mu_i=0.
\label{sf-U}
\end{equation}
with $a_{ij}=(-1)^{x(i)+y(j)} \varphi/4$ forming a staggered-flux
pattern around plaquets of the lattice, as shown in Fig.~\ref{short-fig2}a
(the overall normalization of $U_{ij}$ is of no importance for
the wave function). The gauge-invariant variational
parameter of the staggered-flux ansatz is the flux per plaquet
$\varphi=\sum_\Box a_{ij}$. Even though the ground-state wave
function  breaks time-reversal and translational symmetries,
these symmetries are restored after the projection (\ref{psu2}).

Even with hole doping, the fermion bands are exactly half-filled,
and therefore the number of ``no-fermion'' sites is equal to the number
of ``two-fermion'' sites. This is shown in Fig.\ref{short-fig2}b.  Note
that both these
sites are spin singlets and have the right spin quantum number for a
physical hole.
In $SU(2)$ theory, a $b_1$ boson is attached to the ``no-fermion'' site and
a $b_2$
boson to the ``two-fermion'' site, and both become physical holes, according to
Eq.~(\ref{singlet-states}).
In the mean field theory, the bosons condense to the bottom of their
respective bands, which are located at
${\bf Q}_1 = 0$ and ${\bf Q}_2 = (\pi , \pi)$.  The prescription of
constructing an
$SU(2)$ projected wavefunction is as follows.  The physical wavefunction is
specified
by the location of the up spins and the holes (circled sites in
Fig.\ref{short-fig2}b).  A given set of holes is partitioned into all possible
``no-fermion'' and ``two-fermion'' sites, denoted by $\{ {\bf r}_1 \}$,$\{
{\bf r}_2
\}$.  Each partition specifies a configuration of the staggered flux
wavefunction
given by a product of two Slater determinants (for spin-up and
spin-down spinons).  To this we multiply the phase factor
$
\exp [i\sum_{\{ {\bf r}_1 \} ,\{ {\bf r}_2 \} }( {\bf Q}_1\cdot {\bf r}_1 +
{\bf Q}_2 \cdot  {\bf r}_2)]
$
to represent Bose condensation of $b_1,b_2$ and sum over all partitions.  An
additional sign depending on the ordering of sites is needed to preserve the
antisymmetry of the wavefunction.  This prescription realizes the projection
$P_{SU(2)}$ of a mean-field staggered-flux state to the $SU(2)$ singlet
subspace.


Next we prove the equivalence of the $SU(2)$ projected staggered flux
phase with the pairing state (\ref{wf-pairing}).
Such an equivalence
is known in the undoped case \cite{Zhang}, we extend it to doped
systems.
We note that the mean-field ground state of the staggered flux phase
has the form
$
 |\Phi_{mean} \rangle =
|\Phi_{f} \rangle\otimes
|\Phi_{b} \rangle
$, where
$|\Phi_{b} \rangle = e^{\bar b_1 B_1^\dagger+ \bar b_2
B_2^\dagger}|0\rangle$ with
$(B_1, B_2) = (\sum_i b_{1i},\sum_i (-1)^i b_{2i})$, and
$|\Phi_{f} \rangle$ is the fermion state in the staggered flux phase
Eqs.~(\ref{mf-ham}),(\ref{sf-U}).
Complex numbers $\bar b_1$ and $\bar b_2$
parametrize the Bose condensate where
$b_{1,2}$ condense into their band bottoms:
$B_a|\Phi_b\rangle=\bar b_a |\Phi_b\rangle$.
The $SU(2)$ projection selects the subspace with equal numbers of $B_1$
and $B_2$ bosons, and therefore states corresponding to different
choices of $\bar b_1$ and $\bar b_2$ differ after projection only by
a one-parameter transformation $\exp(\lambda N_h)$, where $\lambda$ is a
complex parameter, $N_h$ is the number of holes. If we fix the average
hole concentration, the wave function is defined unambiguously,
up to an unimportant gauge $\exp(\lambda N_h)$ with ${\rm Re} \lambda=0$.


The staggered-flux mean-field state is
related to the $d$-wave state by a $SU(2)$ rotation
$W_i=\exp[(-1)^i{\pi\over4} \tau_1]$:
\begin{equation}
U^\prime_{ij} =
W_iU_{ij}W_j^\dagger
\propto \pmatrix{ 1 & \pm\Delta/2 \cr \pm\Delta/2 & -1 }
\end{equation}
where the sign
of $\Delta$ is opposite for vertical and horisontal links, and $\Delta$
related to $\varphi$ by 
\begin{equation}
\tan{\varphi\over 4}={\Delta\over 2}.
\end{equation}
After the $SU(2)$ rotation, the mean field state has
the same form, except $|\Phi_f\rangle$ is replaced by
$|\Phi_f^\prime \rangle$, the fermion $d$-wave state of the BCS form
(\ref{wf-pairing}), and the bosonic parameters $\bar b_a$ are rotated:
$\bar b_1^\prime=(\bar b_1+i\bar b_2)/\sqrt{2}$,
$\bar b_2^\prime=(i\bar b_1+\bar b_2)/\sqrt{2}$.
Since the two mean-field states
are related by a $SU(2)$ gauge transformation, they lead to the same
physical state after the $SU(2)$ projection:
$P_{SU(2)}|\Phi_{mean}\rangle
=P_{SU(2)}|\Phi_{mean}^\prime \rangle$.
The freedom in the choice of the bosonic parameters $\bar b_1$ and $\bar b_2$
established in the staggered-flux gauge allows us to set
$\bar b_2^\prime=0$, then the $SU(2)$ projection becomes equivalent
to the conventional Gutzwiller projection.
This proves that the $SU(2)$
wave function is identical to the conventionally-projected wave function.
Of course, the above proof equally applies to the systems with fixed
numbers of holes which we use in our VMC calculation. In this case,
the $SU(2)$ projected wave function is identical to that given by
Eq.~(\ref{wf-pairing}).

Note that in the proof of the equivalence of the two wave functions
we use $a^\mu_i=0$ (and, therefore, $\mu=0$ ). A finite
$\mu$ in the pairing gauge corresponds to a nonzero value of the
Lagrange multiplier $a^3_i$ (chemical potential). In the staggered-flux
gauge, this translates into an on-site pairing term
\begin{equation}
\Delta(\theta_i)=\mu(e^{i\theta_i} f^\dagger_{\uparrow i}
f^\dagger_{\downarrow i} + e^{-i\theta_i} f_{\downarrow i}
f_{\uparrow i})
\end{equation}
with some site-dependent phases $\theta_i$. This term only slightly
affects the properties of the projected wave function, and
we neglect it in the further discussion.

In the ground state of the staggered-flux mean-field Hamiltonian
[eq.(\ref{mf-ham}) with $U_{ij}$ given by (\ref{sf-U})], the $b_1$
and $b_2$ bosons attract each other. This can be understood from the
correlation function for the ``excess density of fermions'':
\begin{eqnarray}
\langle \left( 1-n_f(i) \right) \left( 1-n_f(j) \right) \rangle
& = & \langle \left( 1-f^\dagger_{\alpha i} f_{\alpha i} \right)
\left( 1-f^\dagger_{\beta j} f_{\beta j} \right) \rangle \nonumber \\
& = &
-\big| \langle f^\dagger_{\alpha i} f_{\alpha j} \rangle \big|^2 <0
\end{eqnarray}
at the mean-field level. This means that around a ``no-fermion''
hole ($b_1$ hole) with
$1-n_f(i)>0$ there is a region of an increased probability to find
a ``two-fermion'' hole ($b_2$ hole) with $1-n_f(j)<0$,
and vice versa. The mean-field Green's function
$ G(i,j)=\langle f^\dagger_{\alpha i} f_{\alpha j} \rangle $
decays as $R^{-2}$ at large distances $R=|i-j|$. This is a consequence of the
nodes $k=(\pm\pi/2,\pm\pi/2)$ in  the mean-field spectrum
\begin{equation}
E(k)\propto \sqrt{\cos^2 k_x +\cos^2 k_y + 2 \cos{\varphi\over2}
\cos k_x \cos k_y}.
\end{equation}
Thus at the mean-field level, the attraction of the two species of holes
leads to a $R^{-4}$ decay of the density-density correlations.
After the projection,
the attraction becomes much weaker, but still survives, as we shall
see from our VMC computations.  This attraction between holes was observed
earlier
\cite{Gros}, but was difficult to explain in the $d$-wave gauge.

\begin{figure}
\epsfxsize=3.0in
\centerline{\epsffile{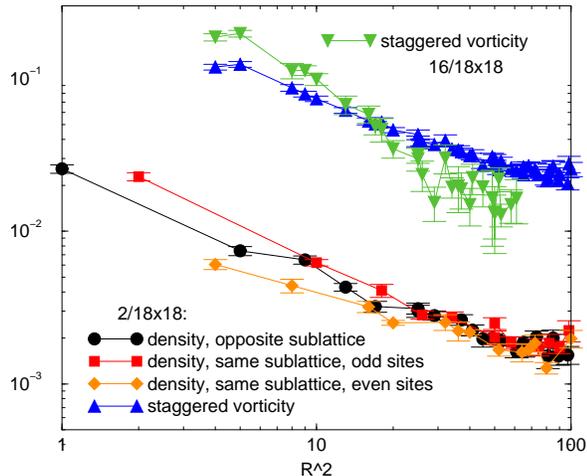}}
\smallskip
\caption{Hole density and staggered-vorticity correlations for 2 holes
and staggered-vorticity correlations for 16 holes in the $18\times 18$ lattice.
Boundary conditions are periodic in one and antiperiodic in the other
direction, $\Delta=0.55$. The correlation functions are
divided by the density of holes $x$ and plotted as a function of the
squared distance. The data are obtained as a result of averaging over
$2 \cdot 10^4$ samples for $2/18\times 18$ system and $2\cdot 10^3$
samples for $16/18\times 18$ system.}
\label{short-fig3}
\end{figure}

In Fig.~\ref{short-fig3} we plot the correlations of the hole
density $n_h(i)=1-c^\dagger_{\alpha i} c_{\alpha i}$ and of the
vorticity $V$ at $\Delta=0.55$
for two holes in a $18\times 18$
lattice as functions of distance.
We observe power-law decay of both correlations,
$\langle n_h(0) n_h(R) \rangle \propto \langle V(0) V(R) \rangle \propto
R^{-\alpha}$, with the equal exponents $\alpha \approx 1.2$.

The reduction of the exponent $\alpha$ from its mean-field value $\alpha=4$
is due to the projection. For the case of two holes (``zero-doping limit''),
this reduction is so strong that $\alpha<2$ and, as a consequence of the
sum rule $\sum_R \langle n_h(0) n_h(R) \rangle = n_h$ , the two holes
become unbound as the size of the system increases.

The proportionalty
of the vorticity and density correlations, together with the sign of the
vorticity correlations, suggests that the two holes may be thought of
as carrying opposite staggered vorticity. 
This can be justified if we fix the gauge and describe the wavefunction as the
staggered-flux phase. At the mean-field level,
this breaks the time-reversal
symmetry, and the fermions in the filled single-particle states carry
a non-zero staggered-vorticity with
$\langle j^f_{ij} \rangle \ne 0$,
where $j^f_{ij}=i (f^\dagger_{\alpha i} f_{\alpha j} -
f^\dagger_{\alpha j} f_{\alpha i})$ is the (unphysical) fermion current.
The $SU(2)$ projection (\ref{psu2}) restores the time-reversal symmetry,
and for the projected wave function
$\langle j_{ij} \rangle = 0$, for the physical current $j_{ij}$.
In the staggered-flux gauge (\ref{sf-U}), at the mean-field level,
$j_{ij}\sim j^f_{ij}$ for $b_1$ holes and $j_{ij}\sim -j^f_{ij}$ for
$b_2$ holes. Thus $\langle j_{ij}\rangle =0$ for the physical current
is a result of the balance between the opposite staggered-vorticity
of $b_1$ and $b_2$ holes.
The attraction between the two $SU(2)$ species of holes then implies
attraction between holes circulating in the opposite directions. 
For a
finite system with two holes, the holes are always of opposite $SU(2)$ types,
which resultins in the proportionality between the vorticity and density
correlations.

The above picture can be summarized in the expression
 $\tilde V \propto \rho_1-\rho_2$, where  $\tilde V$ is the
staggered-vorticity $(-)^R V(R)$ for the physical current
$j_{ij}$, and $\rho_{1,2}$ are the densities of the two bosons
$b_{1,2}$. The correlations of the staggered-vorticity and of 
$ \rho_1-\rho_2 $ are related
$\langle \tilde V(R) \tilde V(0) \rangle 
\propto  
  \langle \rho_1(R) \rho_1(0) \rangle 
 +\langle \rho_2(R) \rho_2(0) \rangle 
-2\langle \rho_1(R) \rho_2(0) \rangle  
$.
When there are only two holes $\langle \rho_1(R) \rho_1(0) \rangle = 
\langle \rho_2(R) \rho_2(0) \rangle=0$, we find that 
\begin{equation}
\langle \tilde V(R) \tilde V(0) \rangle \propto 
- \langle \rho(R) \rho(0) \rangle  
\label{tVrho}
\end{equation}
where $\rho=\rho_1+\rho_2$ is the total density of the holes. 
The minus sign explains the $\pi$ phase shift seen in Fig.~1.




If we
assume pairing between holes of opposite staggered vorticity, we may
interpret the
vorticity correlations as the hole correlations within one pair which
allow us to determine the strength of the pairing correlation (even for finite
density of holes) and measure the size
of pairs. At finite density of holes, the correlation of the
staggered-vorticity decay faster. In Fig. 4b we present the correlation for
16 holes in $18\times 18$ lattice (nearly 5\% doping) at $\Delta=0.55$. We
find $\alpha\approx 2.2$. The increase of $\alpha$ above $2$ may be
interpreted as a formation of bound hole pairs (and hence the onset of
superconductivity).
The small value of $\alpha-2$ after the projection
implies that the pairs are bound very loosely and their size is large.
In this case, the nearest-neighbor pairing amplitude
$\langle c_{i\uparrow}c_{j\downarrow} \rangle$ is small, which may account for
numerical conclusions about the absence of superconductivity at
$J/t<0.5$
\cite{Shih,Heeb}.

While the staggered vorticity correlation is found for a superconducting
wavefunction,
we speculate that the phase coherence of the bosons may not be crucial and
that such
correlation may survive above T$_c$ in the pseudogap state, and indeed
serves as a
signature of that state.  Unfortunately, detection of this correlation may be
difficult.  We estimate that the fluctuating current generates a
fluctuating staggered
magnetic field of order 40~G.  This field will contribute to the relaxation
of the $Y$
nuclei, which are ideally sited above the center of the plaquets.
However, we do
not have dynamical information at present and it is difficult to predict
the magnitude
of this orbital relaxation.  Another possibility is to freeze in some
staggered field
pattern around an impurity site.  If some $Y$ is replaced by an impurity
with spin
${\bf S}$ which strongly couples to the Cu-O plane, we may use a magnetic
field to
align the spin and therefore create a static orbital current via the
${\bf L}\cdot {\bf S}$ term. This local orbital current may then generate a
static
staggered pattern around it, which may be detected by shifts in the $Y$ NMR
line.
Perhaps a more promising way to test our prediction is to look for this
effect in
exact diagonalization of small systems.

We thank H. Alloul for helpful discussions and we acknowledge the support
by NSF under
the MRSEC program DMR98-08941.

\end{document}